\documentclass{article}

\usepackage{arxiv}

\usepackage[utf8]{inputenc} 
\usepackage[T1]{fontenc}    
\usepackage{hyperref}       
\usepackage{url}            
\usepackage{booktabs}       
\usepackage{amsfonts}       
\usepackage{amsmath}
\usepackage{amssymb}
\usepackage{nicefrac}       
\usepackage{microtype}      
\usepackage{graphicx}
\usepackage{natbib}
\usepackage{doi}
\usepackage{orcidlink}
\usepackage{multirow}

\title{Towards triggerless four-dimensional detectors for High Energy Physics collider experiments}

\author{Vladimir V. Gligorov\orcidlink{0000-0002-8189-8267}\\
LPNHE, Sorbonne Universit{\'e} \\
Paris Diderot Sorbonne Paris Cit{\'e} \\ 
CNRS/IN2P3, France \\
\href{mailto:vladimir.gligorov@cern.ch}{vladimir.gligorov@cern.ch}}

\renewcommand{\headeright}{}
\renewcommand{\undertitle}{}
\renewcommand{\shorttitle}{Review of real-time data processing for collider experiments}

\hypersetup{
pdftitle={A template for the arxiv style},
pdfsubject={q-bio.NC, q-bio.QM},
pdfauthor={David S.~Hippocampus, Elias D.~Striatum},
pdfkeywords={First keyword, Second keyword, More},
}

\begin{document}
\maketitle
\begin{abstract}
High Energy Physics (HEP) experiments at flagship colliders produce and process some of the biggest datasets on Earth, with the current generation of flagship experiments at the Large Hadron Collider (LHC) producing more than a tenth of the world’s total internet traffic every second. Moreover the quantities of data produced have increased exponentially over the past decades and this trend shows no sign of slowing down. In parallel, the use of picosecond timing is becoming more common in HEP detectors, enabling qualitatively new approaches to real-time processing and selections. I review the planned introduction of precision timing information into the upcoming upgrades of the CMS, ATLAS, and LHCb experiments. I discuss the ways in which the combination of timing and networking technology may enable future detectors to be designed as triggereless from the ground up, and reflect on the physics benefits of such a paradigm shift for the field. 
\end{abstract}

\newpage
\section{Introduction} 

\begin{figure}[t]
    \centering
    \includegraphics[width=0.7\textwidth]{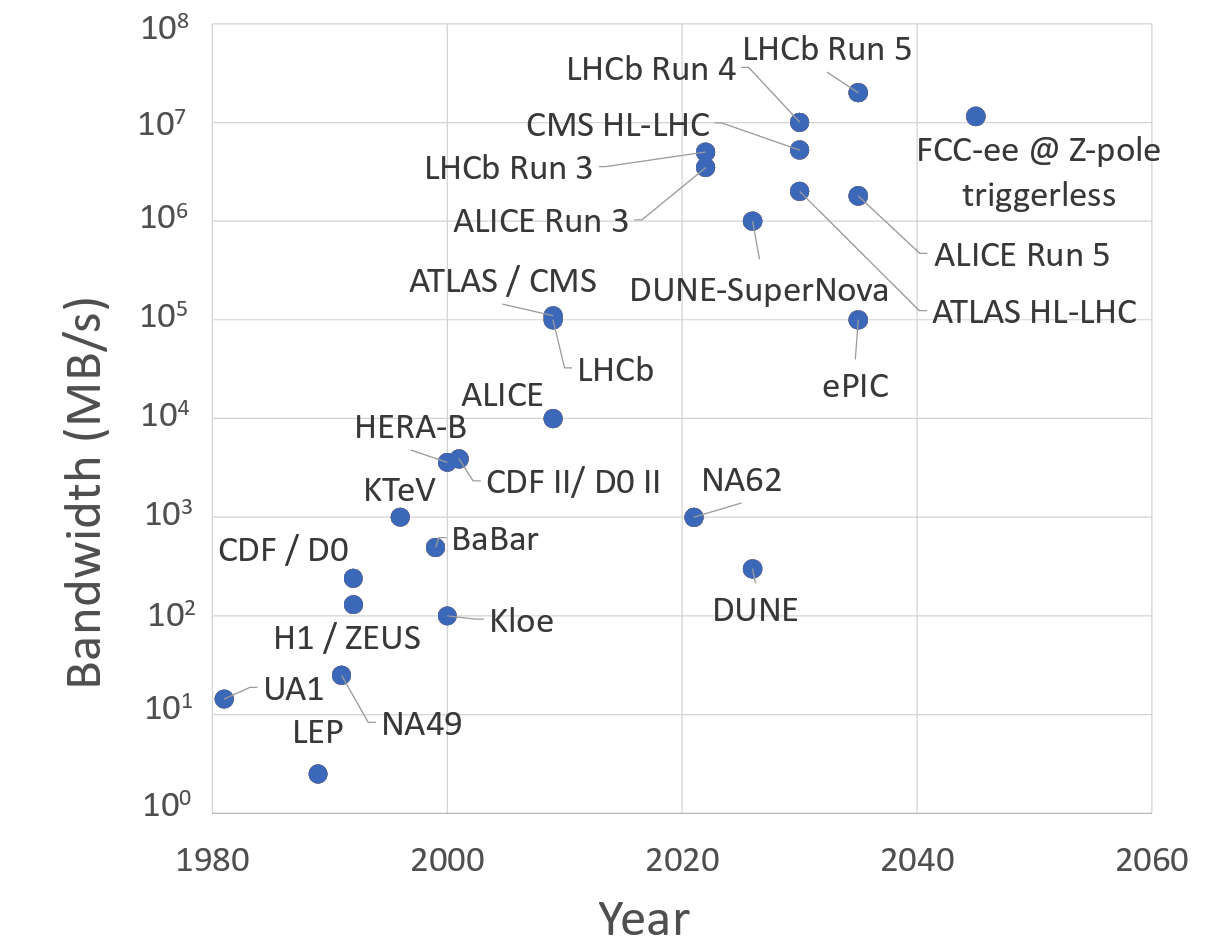}
    \caption{Instantaneous data rates (bandwidth) of HEP experiments over the past four decades. 
    Data compiled by A. Cerri and D. vom Bruch and reproduced from~\cite{deBlas:2025gyz} as 
    permitted by the article's licence.}
    \label{fig:data_rates_cerri}
\end{figure}

High Energy Physics (HEP) experiments at flagship colliders produce and process some of the biggest 
datasets on Earth, with the current generation of flagship experiments at the Large Hadron Collider (LHC) 
producing more than a tenth of the world’s total internet traffic every second. Moreover the quantities 
of data produced have increased exponentially over the past decades, as shown in 
Figure~\ref{fig:data_rates_cerri}. Because it is prohibitive to store such data volumes, 
much less transmit them to individual physicists for analysis, experiments must reduce them by orders 
of magnitude in real time, storing only the fragments which are most of interest to physics analysis. 

Two constraints have traditionally governed the design of this real-time data processing. First of all, 
it is prohibitively expensive to perform a physics-analysis-quality processing of the full data rate 
collected by HEP experiments. This cost is typically factors larger than the cost of the experiment as
a whole, before considering the fact that a physics-analysis-quality processing requires data to be 
buffered while the detector alignment and calibrations which must be applied to this data are computed. 
Secondly, the cost to even transfer the full data rate off the detector has typically been prohibitive.
This cost is financial, because of the number of data links required, as well as physical, because in
hermetic detectors the data links for the innermost detector components must necessarily lie in the
detector's geometric acceptance, increasing the material budget and therefore worsening the detector's
resolution and physics performance.

For these reasons, the necessary real-time data reduction was traditionally carried out by “trigger” 
systems, which reconstruct a subset of the detector information and use it to identify collider bunch 
crossings (``events'') which are most likely to contain interesting physics processes. 
Although trigger systems have intrinsically 
worse resolutions and efficiencies than the physics-analysis-quality reconstruction, they require far 
less computing power and data bandwidth, making them historically indispensable in real-time processing. 
Most modern HEP trigger systems are based on the fundamental principles introduced by the UA1
trigger~\cite{Dorenbosch:1985cx}, and in particular on the idea of of a processing cascade in which
each trigger stage performs a more complex and complete processing of the reduced data rate selected
by the previous stage. Because both collider experiment data rates and storage costs~\cite{prices} have 
evolved exponentially over the past decades, one increasing while the other decreased, triggers tended
to reduce data volumes by around four orders of magnitude almost independently of collider type or 
experiment. 

Data collected by a trigger system typically undergoes further ``offline'' processing. The best 
available detector alignment and calibrations are applied to the data which is then reconstructed 
with the most efficient algorithms available. Subsequently, high-level physics objects used by physics
analysis are formed from the reconstruction output and serve as inputs to the final physics analysis.
These high-level physics objects themselves typically take one or two orders of magnitude less storage
space than the event which they were found. The extra information carried by these
objects compared to the objects available to the trigger system, and the improved resolution on their
properties made possible by the analysis-quality reconstruction, calibration, and alignment, allows
a further one to two order of magnitude reduction in data volumes by applying offline selections to 
reduce backgrounds. In total, therefore, a further reduction of roughly two to four orders of magnitude
in the required storage space, on top of what the trigger already achieved, is possible once
the full physics-analysis-quality data processing can be used.

\begin{figure}[t]
    \centering
    \includegraphics[width=0.7\textwidth]{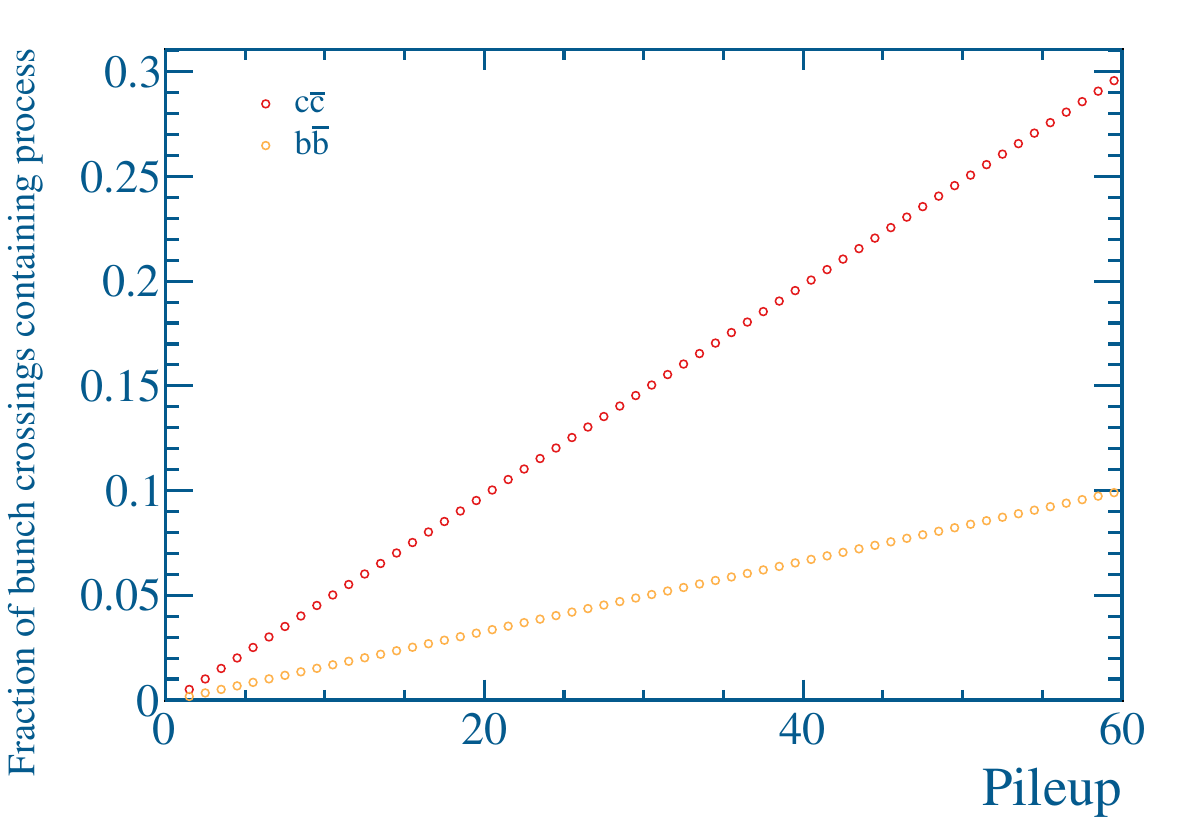}
    \caption{The fraction of LHC bunch crossings which produce a $b\bar{b}$ or $c\bar{c}$ quark pair that can be partially reconstructed in the 
    acceptance of the LHCb detector as a function of the number of proton-proton collisions in a single
bunch crossing (``pileup''). Based on studies originally documented in~\cite{Fitzpatrick:1670985}.}
    \label{fig:lhcbbbarvspileup}
\end{figure}

Because triggers reduce the data volume by selecting entire events, they are only useful if the
physics of interest is rarely produced and if the associated backgrounds can be significantly reduced
using only partial detector information. In practice this requires the processes of interest to have
production cross-sections many orders of magnitude smaller than the total inelastic cross-section
at the relevant collider. This requirement began to break down at the Large Hadron Collider as the
instantaneous luminosity was increased by packing more and more proton-proton collisions in a single
bunch crossing. This is illustrated in Figure~\ref{fig:lhcbbbarvspileup} for the LHCb detector, which
operated with a pileup of $\approx\! 1.1$ between 2011 and 2018 (Runs~1~and~2 of the LHC) and with a 
pileup of $\approx\! 5.3$ between 2024 and today (Run~3 of the LHC), and is planning to operate with a 
pileup of $\approx\! 25$ after 2036 (Run~5 of the LHC). At a pileup of $\approx\! 1$ it is possible to 
select events in which a $b\bar{b}$ pair was produced using a traditional trigger, but it is  difficult 
to use the same approach for $c\bar{c}$ pairs which are already produced in $\approx\! 0.5\%$ of events. 
By the time pileup is $\approx\! 5$ traditional triggering is difficult even for $b\bar{b}$ pairs, 
and by the time pileup is $\approx\! 25$ this approach breaks down entirely. 

Similar challenges were faced by other LHC experiments in their
respective contexts. For example when searching for low-mass dark matter particles predominantly 
decaying into jets, the irreducible Standard Model QCD backgrounds rapidly saturate the trigger system.
These difficulties were compounded by the exceptional performance of the LHC detectors themselves, 
which proved capable of precision physics measurements far beyond their initial scope. Together with
the now roughly thirty-year planned runtime of the LHC, it became crucial to develop new approaches
to real-time data processing in order to maximise the scientific exploitation of these once in a 
generation instruments.

The reduction in data volume which can be achieved offline once analysis-quality information is
available offers a way out: if such information can be made available in real-time, then instead
of recording the full detector information for a small subset of events, a trigger system can 
selectively record only those high-level physics objects actually required for analysis for
a much larger subset of events. Over the past ten years, all four major LHC experiments (ALICE, 
ATLAS, CMS, and LHCb) have worked in this direction and demonstrated~\cite{Khachatryan:2016ecr,
Aaij:2016rxn,ATL-DAQ-PUB-2017-003,ATLAS:2018qto,ALICEO2} the ability to perform 
physics-analysis-quality detector reconstructions in real or quasi-real time, as 
well as to calibrate their detectors on timescales compatible with real-time processing constraints. 
These developments, which I will collectively refer to as “real-time analysis”, have mostly lifted 
the first of the two aforementioned constraints on the design of real-time data processing. 
While none of the experiments is currently performing 
a physics-analysis-quality processing of their full data rate in real time, all of them developed ways
to create fully aligned and calibrated high level physics objects which can be directly used in 
analysis in real time for at least a subset of physics analyses. And in the case of LHCb and ALICE, 
the full physics-analysis-quality detector reconstruction can now be performed on O($10\%$) of the 
data rate, at least two orders of magnitude above what was possible for the previous generation of 
experiments. 

The second constraint, of the bandwidth required to transfer the full detector data into near-detector 
processing centres, has already been lifted for non-hermetic detectors such as LHCb, where the material
budget penalty associated with the readout cabling does not apply. It seems plausible that technological
developments will fully lift this constraint for hermetic detectors in the coming decades.
It is therefore timely to start conceptualising how the next generation of flagship collider experiments, 
in particular at post-LHC facilities, might be built around triggerless real-time analysis and benefit from it.
Such concepts must also confront the increasing availability of precision time information in collider 
experiments, which simultaneously increases data rates and adds a powerful handle for differentiating 
between interesting physical processes and backgrounds. This review will discuss current trends in detector
design, networking, and data processing and advocate that these are converging so as to make a full
real-time analysis-quality four-dimensional detector processing the natural solution for the next generation
of collider experiments. More details on the evolution of collider
trigger systems and real-time analysis over the past decades can be found in~\cite{Gligorov:2023ezo}. 
A recent review of the technology behind timing detectors and its evolution can be found 
in ~\cite{MALBERTI20261} and is complementary to this more physics-oriented article. 
Natural units $\hbar = c = 1$ are used throughout this paper.

\section{From dedicated timing measurements to four-dimensional detectors}
All HEP detector electronics needs to be able to precisely timestamp the signals which it records
in order to associate hits in the different parts of a detector with the correct event. At the LHC
the bunches are spaced 25~ns apart, so timestamping of $O(1)$~ns is already sufficient for this purpose.
On the other hand the individual $pp$ collisions within a given LHC event are distributed~\cite{Bruning:782076} 
with an RMS of $\approx\! 200$~ps, so associating individual particles with the $pp$ collision in which
they originated requires a time resolution of $O(10)$~ps per particle. Finally the difference in flight 
times for a Kaon and a pion at 10~GeV$/c$ momentum is roughly 3.75~ps per metre, so time-of-flight (TOF)
particle identification detectors operating at LHC energies require either femtosecond-level timing 
or a way of increasing the distance travelled before the signal is recorded. While the first of these
three uses of timing does not give any information about the properties of individual particles produced 
within an event, the subsequent two certainly do. For the purposes of this review a 
four-dimensional detector will therefore be taken to mean a detector in which the recorded hit times 
are roughly as important to reconstructing the properties of individual particles as the recorded hit 
space coordinates. 

An important property of precision timing information is that it is subject to different and generally
uncorrelated systematic uncertainties and interpretation biases compared to precision spatial information.
We will first discuss the use of precision timing information in existing, including non-collider or low-energy,
experiments, before turning to the planned use of precision timing information in the ATLAS, CMS, and LHCb 
detectors during the high-luminosity LHC (HL-LHC) period. Charged particle trajectories are conventionally
referred to as ``tracks'' throughout.

\subsection{Use of timing for low-energy particle identification}

Time-of-flight detectors have been used for charged particle identification across a wide range of HEP
detectors. A recent representative example is the ALICE TOF detector based on multigap resistive
plate chambers, which covers the central rapidity region of roughly $-0.9 < \eta < 0.9$ and whose performance 
is summarised in~\cite{Carnesecchi:2018oss}. Its essential figure
of merit is the per-track time resolution, which with the most recent calibrations reaches 56~ps, enabling
pion-kaon separation below 3~GeV of momentum and kaon-proton separation below 5~GeV of momentum. In
order to benefit from this intrinsic detector resolution, ALICE needs to determine the time of the 
collision\footnote{ALICE refers to the collision time as ``event time'' in its publications, but I have chosen
to use collision time here in order to have a unified nomenclature with respect to experiments which must
timestamp multiple collisions per bunch crossing.} which produced the particles and to correct for the energy 
lost as the particles traverse the ALICE detector material. The time of the collision is simplified compared 
to other collider experiments by the fact that ALICE almost always operates at an instantaneous luminosity 
corresponding to one or fewer collisions per bunch crossing. 

\begin{figure}[t]
    \centering
    \includegraphics[width=0.7\linewidth]{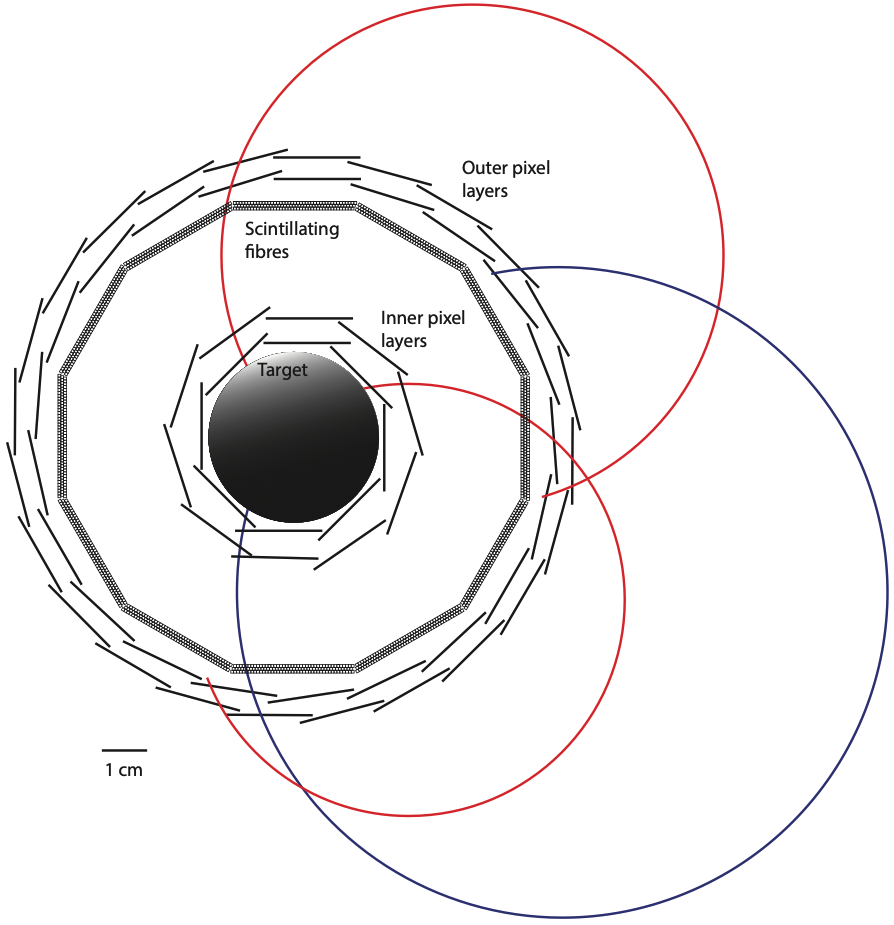}
    \caption{Layout of the Mu3e detector, reproduced from~\cite{Mu3e:2020gyw} as permitted by the article's licence.}
    \label{fig:mu3elayout}
\end{figure}

While ALICE has two specialized high-rapidity detectors for measuring the collision time, their 
narrow angular coverage makes them of little relevance to the global detector reconstruction. The 
technique~\cite{ALICE:2016ovj} used to derive the collision time from the TOF detector information 
alone is, on the other hand, more generally applicable. Each track which can be associated to TOF
detector information via some quality criteria provides an estimate of the collision time. This
estimate $t_{coll}$ is given by the difference between the track time in the TOF detector and the expected
time of flight between the collision point and the TOF detector for a particle of a given mass 
and momentum. The uncertainty on this estimate is given by the quadratic sum of the intrinsic TOF
detector resolution and the uncertainty due to the track reconstruction, which depends on the track
momentum and the particle species (and therefore mass) assigned to the track. For any given
set of assumed particle species, the collision time can be estimated as
\begin{equation}
t_{coll} = \sum_{tracks}(w_i\cdot t^{i}_{coll})/\sum_{tracks}(w_i),
\end{equation}
where $w_i$ is a per-track weight given by the inverse square of the estimated uncertainty on 
the per-track event time estimate $t^{i}_{coll}$. 

Since the particles cannot be identified before the collision time is known, the estimated time must
be calculated by varying the assigned particle species in order to minimize the spread between each 
track's estimate of the collision time and the weighted average estimate of the collision time
\begin{equation}
    \chi^2 = \sum_{tracks}((t^{i}_{coll}-t_{coll})/\sigma^i_{t_{coll}})
\end{equation}

This estimated collision time is, however, biased with respect to each individual track's assumed
particle hypothesis. In principle, a separate collision time should be computed for each track using
only information given by \textit{other} tracks with TOF information. In order to balance physics 
accuracy and computational time, ALICE chooses to group tracks in ten bins of momentum and compute 
the collision time for each bin using only tracks found in the other bins. This is analogous to the 
problem of computing a decaying long-lived particle's flight distance with respect to its production 
point (vertex), in which tracks from the long-lived particle decay can bias the estimate of the production point position 
if they are not removed from the production vertex fit. In general, although the ALICE TOF detector is limited 
in scope, particularly in terms of its momentum coverage, we will see that the basic principles of 
its reconstruction generalise well to precision timing applications in other collider experiments.

\subsection{Use of timing in the Mu3e experiment}

The Mu3e experiment~\cite{Mu3e:2020gyw} is designed to search for the $\mu^+\to e^+ e^- e^+$ decay, 
which is essentially forbidden within the Standard Model with a branching fraction of $O(10^{-55})$ 
from diagrams involving neutrino oscillations. Probing branching fractions of $O(10^{-16})$ requires 
Gigahertz rates of stopped muons, and a detector which can accurately separate genuine 
$\mu^+\to e^+ e^- e^+$ decays from the dominant combinatorial backgrounds. A cross-section layout 
of the Mu3e detector is shown in Fig.~\ref{fig:mu3elayout}.

Mu3e will use two types of timing detectors: a thin scintillating fibre tracker with a target time resolution
of 250~ps placed around the target region, and thicker tile detectors with a target time resolution of below 
100~ps placed forward and backwards of the target region. The experiment will be read out in 50~ns
time slices, each of which will contain~\cite{vomBruch:2017khm} around 10 tracks from Standard Model processes 
in the first phase of the experiment and around 100 tracks in the second, more sensitive, phase.

By comparison to the ALICE TOF detector, there is no such thing as a collision time in Mu3e, but all genuine 
tracks in the detector are known to be electrons. The timing information is primarily used to evaluate the 
coincidence between the three electrons forming a signal candidate and thus reject backgrounds in which
multiple stopped muons decay close in space, but far away in time, within any given 50~ns time window. 
A secondary use of the timing information is to suppress fake tracks and to identify tracks with ambiguous
charge assignments. Depending on its trajectory a track will either have two scintillating fibre hits or
one scintillating fibre and one tile detector hits. The compatibility between the time difference of these 
two hits and the track trajectory length allows both fake segment combinations and wrong charge assignments 
to be suppressed. The use of timing coincidence information for the suppression of combinatorial backgrounds
is also relevant for collider experiments seeking to resolve multiple collisions per bunch crossing.

\subsection{Use of timing in the NA62 experiment}

\begin{figure}[t]
    \centering
    \includegraphics[width=0.9\linewidth]{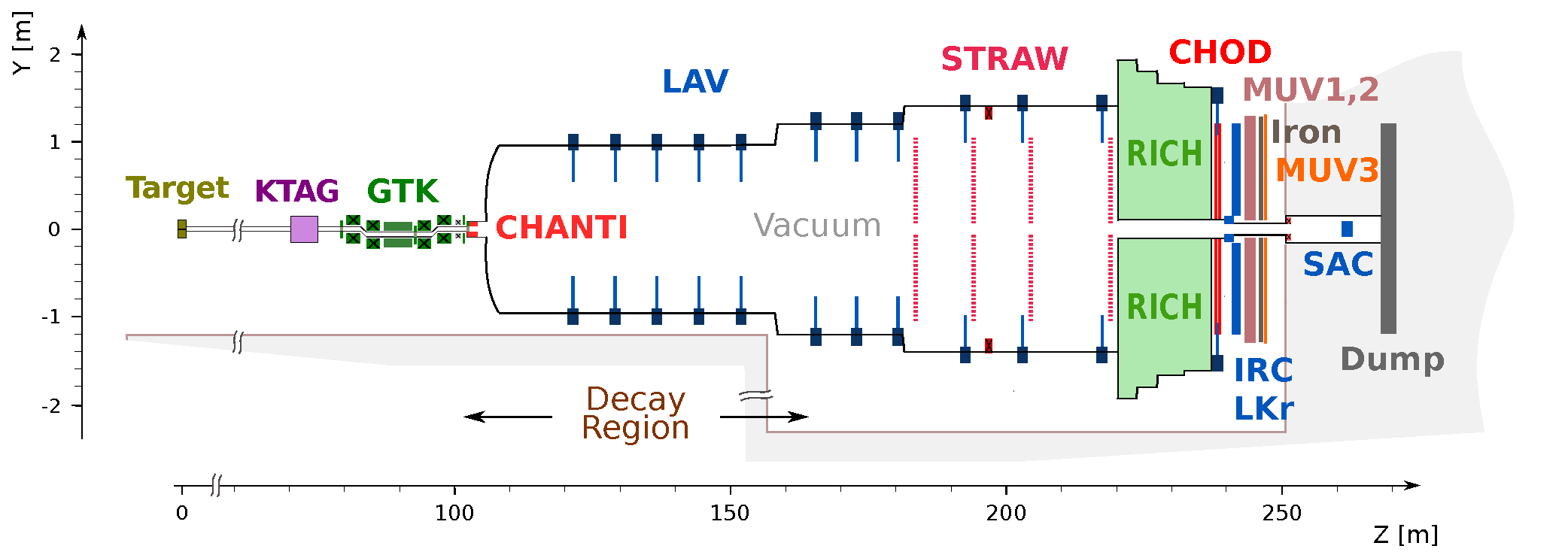}
    \caption{Layout of the NA62 detector, reproduced from~\cite{NA62:2017rwk} as permitted by the article's licence.}
    \label{fig:na62layout}
\end{figure}

The primary physics goal of the NA62 experiment~\cite{NA62:2017rwk} is the precise measurement of the
$K^+\to \pi^+\nu\bar{\nu}$ branching fraction, which is predicted to be $O(10^{-10})$ in the Standard Model
but can be enhanced or further suppressed in a wide range of beyond Standard Model theories. Similarly
to Mu3e, the experiment requires a large incoming flux of parent particles, in this case 45~MHz of kaons, 
as well as an excellent suppression of Standard Model backgrounds which are, contrary to Mu3e, abundant.
The layout of the NA62 detector is shown in Figure~\ref{fig:na62layout}.

The basic physics signature of the $K^+\to \pi^+\nu\bar{\nu}$ decay is a reconstructed track which
is positively identified as a kaon and is matched to an outgoing track which is positively identified
as a pion, with no other detector activity. Precision timing information for tracks is provided by the
\textbf{KTAG} Cherenkov counter which positively identifies incoming kaons with a time resolution of 
70~ps, the \textbf{GTK} pixel tracker which precisely measures the incoming kaon momentum and 
direction and has a per-pixel time resolution of 150~ps, a \textbf{RICH} Cherenkov detector
which separates pion and muon tracks while measuring the track crossing time with a resolution of
100~ps, and a \textbf{CHOD} system of hodoscope counters with time resolution of around 150~ps.
Because the spread of the incoming kaon momenta is significant relative to the required kinematic 
separation between the signal $K^+\to \pi^+\nu\bar{\nu}$ and background $K^+\to \pi^+\pi^0$ processes,
the time coincidence of signal candidates in these detectors is essential for suppressing combinatorial
mismatches between incoming kaon candidates and outgoing pion candidates. Other NA62 detectors 
are also able to timestamp signals, typically with few nanosecond resolutions, which allows to
veto a range of background processes whose reconstructed kinematics fall within the signal region. 
The overall impact of NA62's timing measurements on the resolution with which the key physics property
of the signal decay -- the kinematic relationship between the incoming and outgoing particles -- is
comparable to the impact of the detector's spatial and kinematic resolutions, making NA62 a truly
four-dimensional (though non-collider-based) detector.

\subsection{Use of timing to associate tracks to their origin collisions}

Having considered the use of timing information in several experimental scenarios, we can now turn to
the one of the most common and general problems facing detectors at high-luminosity colliders: that 
of associating a reconstructed track to a specific origin collision. This association is essential
for reducing combinatorial backgrounds at high luminosities and is one of the key design requirements
for all upcoming HL-LHC experiments.

\begin{figure}[t]
    \centering
    \includegraphics[width=0.98\textwidth]{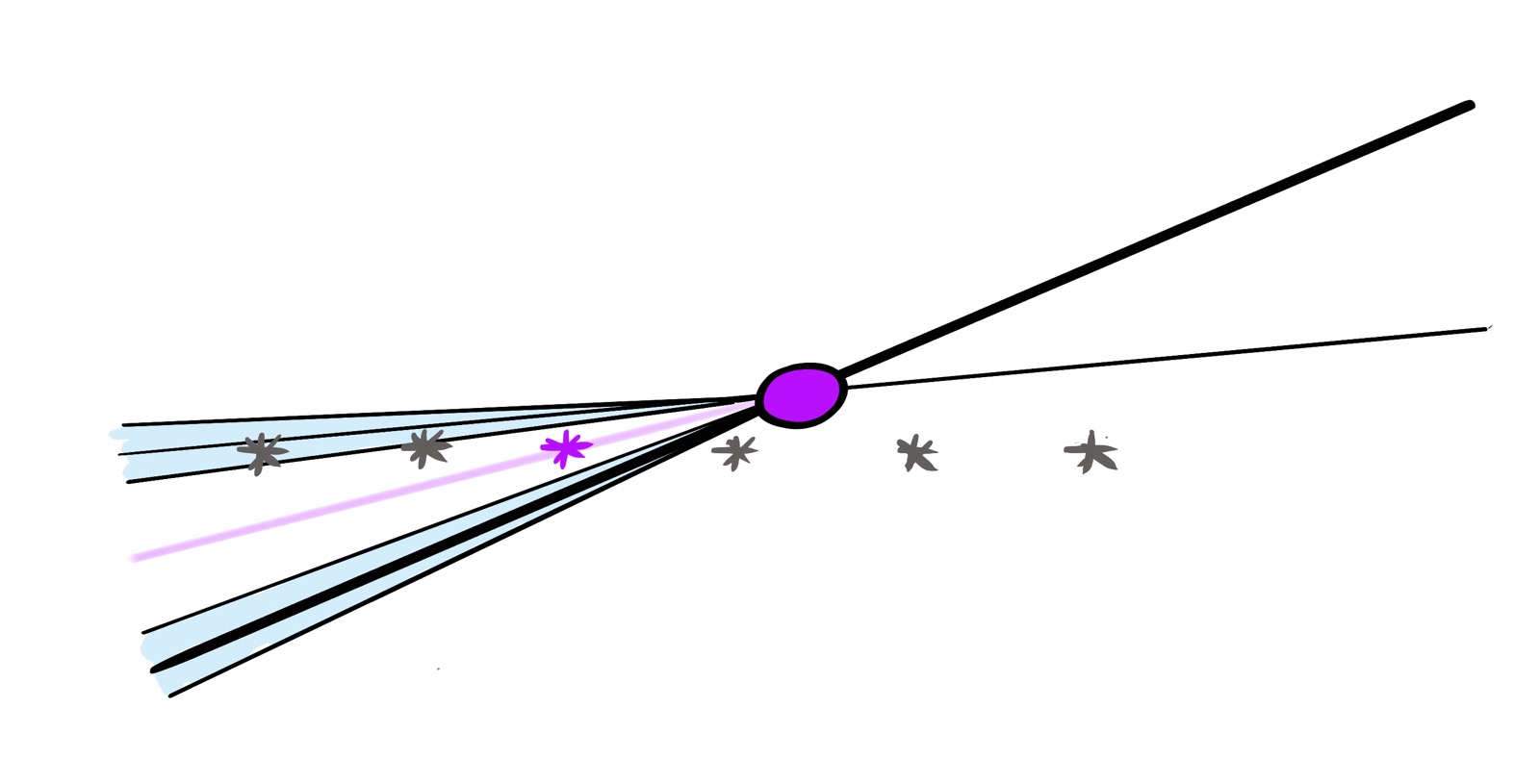}
    \caption{Association of tracks to their origin collision using spatial information. 
    This association is typically based on the distance of closest approach between the track's momentum 
    vector and the collision, otherwise known as the ``impact parameter'' (IP). Tracks produced directly in a given
    collision have an average IP of zero, with a spread given by the IP resolution. Tracks which are produced in
    the decay of long-lived particles have an average IP of zero with respect to their parent particle's collision
    of origin, but the spread of their IP now depends on both the IP resolution and the parent particle's decay time. 
    The one-$\sigma$ band corresponding to the track's IP resolution is shaded blue, the true collision 
    of origin and the decaying long-lived particle are purple, and pileup collisions are grey. The thicker track 
    segment is correctly associated to its parent particle's collision of origin, while the thinner track segment is
    incorrectly associated to a pileup collision. 
    Illustration by Y. Amhis, reproduced from~\cite{amhis_2026_20479510} with permission.}
    \label{fig:associating_particles_to_pileup}
\end{figure}

Figure~\ref{fig:associating_particles_to_pileup} illustrates the limitations of this association when 
performed using only spatial information. The smaller the particle momentum, and the further from the 
original collision that particle is created, the higher the probability that this particle will 
be incorrectly associated. This has implications for
physics analysis. For example, requiring all products of a long-lived particle decay to have the
same associated collision results in an inefficiency which depends both on the particle decay-time 
and the number of collisions in a given bunch crossing. This in turn both reduces physics
sensitivity and introduces systematic uncertainties associated with the misassociation modelling. 

If the same association is performed using precision timing information, however, the resolution 
with which the track time can be reconstructed mainly depends on the intrinsic detector resolution and
is therefore largely independent\footnote{This holds for tracks of known momentum, and for relativistic 
tracks generally whether their precise momentum is known or not. These two categories encompass most 
of the tracks that require associating in the first place.} 
of how far from the origin collision the track in question was created. This in turn makes the association 
efficiency and misassociation rate independent of whether the track was produced directly in the original
collision or in the subsequent decay of a long-lived particle.

What this means in practice is that even if the association computed using timing information is 
significantly less precise than that computed using spatial information, using the two together can 
remove the bulk of the bias which arises from using spatial information alone. Consider for example 
a spatial association which is 95\% accurate and a time association which is 80\% accurate. If 
associating a four-body particle decay via spatial information alone, $18.6\%$ of the decays will
have at least one incorrectly associated particle. If however we use the time association as a tiebreaker
in cases where three of the four decay products are associated to the same origin collision, we can
recover around $13.7\%$ of the misassociated decays by simply checking if the fourth particle is
associated to the same collision as the other three using time information. More sophisticated
strategies would of course give better results, but this simple example is hopefully already sufficient
to convince that timing information can not only improve performance but also enable qualitatively different
approaches to data processing and analysis. And in the case of neutral particles, precision timing information
is generally the only way to associate them to their origin collisions.

\subsection{Physics case studies for the use of timing at the HL-LHC}
When discussing timing at the HL-LHC, it is useful to focus on those detectors\footnote{While ALICE also 
plans a second upgrade for Run~5 of the HL-LHC period, its pileup will not significantly change.~\cite{ALICE:2022wwr}} 
which will be operating with a significantly higher pileup than in the past: ATLAS, CMS, and the proposed 
LHCb U2.~\cite{CERN-LHCC-2015-020, Butler:2055167, LHCbcollaboration:2903094}  
Each of these detectors will introduce precision timing information in one or more of their subsystems,
with the principal aim of reducing pileup-induced backgrounds in both the low-level reconstruction and the 
high-level signal candidate selection. The list of the detectors in question and their timing resolutions 
is given in Tab.~\ref{tab:hllhctimingdetectors}.  While the proposed second upgrade of the LHCb detector
would only be installed in the second period of HL-LHC datataking, after 2036, we will for simplicity's sake
refer to all three as HL-LHC detectors in the following discussion.

\begin{table}[t]
    \centering
    \begin{tabular}{l|c|c|c}
         Experiment & Detector & Hit resolution (ps) & Particle resolution (ps) \\
         \hline
         \multirow{4}*{LHCb} &  Vertex detector (VELO) & 50 & <20\\
                             &  Cherenkov detectors (RICH) & 35-100 & <25\\  
                             &  Scintillating fibre tracker (SciFi) (?) & <1000 & <300\\
                             &  Electromagnetic calorimeter (ECAL) & & < 44\\
         \hline
         ATLAS & High-Granularity timing detector (HGTD) & <70 & <50\\
         \hline
         \multirow{4}*{CMS} & MIP barrel timing detector (MTD-BTL) & & <60\\
                            & MIP endcap timing detector (MTD-ETL) & & <40\\
                            & \multirow{3}*{Barrel calorimeter (BCAL)} & & <30 ($E>60$~GeV)\\
                            & & & <150 ($E>10$~GeV)\\
                            & & & <500 ($E>3$~GeV)\\
                            & \multirow{2}*{High-granularity calorimeter (HGCAL)} & \multirow{2}*{<150} & <12 (photons)\\
                            &                                                     &                     & <40 (hadrons)\\
    \end{tabular}
    \caption{Subsystems providing precision timing information for each of the ATLAS~\cite{CERN-LHCC-2020-007}, 
    CMS~\cite{CMS:2667167,CERN-LHCC-2017-023, CERN-LHCC-2017-011}, and 
    LHCb~\cite{LHCb:2021glh,LHCbcollaboration:2903094} HL-LHC detectors. Systems which are not certain to provide
    precision timing information are labelled with a question mark. Where technology choices have not yet been made,
    resolutions are provided as ranges. In the case of ATLAS and CMS subsystems, the quoted estimates are design
    targets at the detector's end of life.}
    \label{tab:hllhctimingdetectors}
\end{table}

\subsubsection{Timing use-cases in ATLAS}
Of the three high-pileup HL-LHC detectors, ATLAS plans to make the most limited use of
timing information. This choice is justified by ATLAS's sufficient tracking and calorimeter 
granularity in the barrel region, such that particles can be associated to pileup collisions 
even without the need for timing. As a result, the HGTD only covers the forward 
pseudorapidity range of $2.4 < \eta < 4.0$. Charged particles with sufficient 
transverse momentum (typically above a few 100~MeV) to be reconstructed in this 
$\eta$ range generally also have high enough momenta that the HGTD cannot be used for 
time-of-flight based particle identification. The primary physics use-case for the HGTD
is therefore to associate charged particles in the forward direction to $pp$ collisions,
and to estimate the time of the collision which is associated with a given signal candidate.

\begin{figure}[t]
    \centering
    \includegraphics[width=0.9\linewidth]{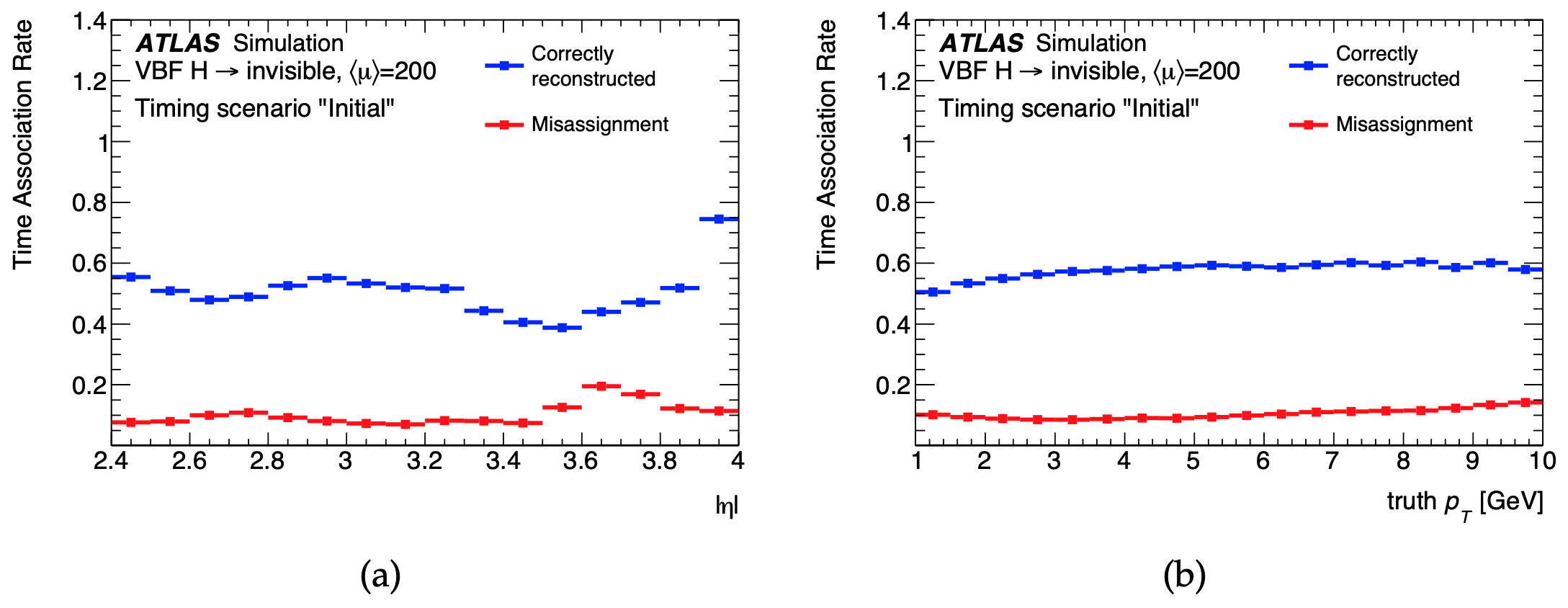}
    \caption{Rate for correct and incorrect assignment of track-times as a function of track $\eta$ (left) and $p_T$ (right). 
    The sum of the two rates gives the inclusive efficiency of track-time assignment. 
    Caption and figure reproduced from~\cite{CERN-LHCC-2020-007} as permitted by the article's licence.}
    \label{fig:hgtd-perf}
\end{figure}

The main inefficiencies in the HGTD come from tracks which undergo material interactions 
before reaching the HGTD detector layers, and which are therefore inaccurately extrapolated
to the HGTD. Such tracks either do not get any hit assigned or get a hit from another track 
with the wrong timing assigned. While material interaction inefficiencies are negligible 
for muons, they reach around 30\% for pions and other hadrons, with an additional 10\%
of tracks having the wrong timing hits assigned. As shown in Figure~\ref{fig:hgtd-perf}, this 
effect is roughly independent of the track kinematics, while variations in $\eta$ are caused by
the detector layout and the distribution of material prior to the HGTD. Because the inefficiencies 
are caused by material interactions rather than detector occupancy, the HGTD is almost fully 
efficient for muons. 

Because of the HGTD's limited forward acceptance and these reconstruction inefficiencies, 
not all $pp$ collisions can be assigned a time estimate. As an example, only around two thirds 
of collisions which produce a Higgs boson via vector-boson fusion also produce a sufficient number
of tracks in the HGTD acceptance for a collision time estimate to be computed. Despite these limitations, 
the HGTD information significantly improves reconstruction and physics object performance across 
a wide range of problems: lepton isolation, the suppression of pileup jets, and the removal of 
calorimeter deposits associated to pileup tracks. In addition, the HGTD can be read out in limited 
form at 40~MHz in order that it can provide a standalone measurement of the luminosity, although 
the HGTD's excellent spatial granularity is more relevant to this task than its timing
performance. 

\subsubsection{Timing use-cases in CMS}

By contrast to ATLAS, the CMS experiment has chosen to introduce precision timing in some form 
throughout its geometric acceptance, leading to a wider range of physics use-cases and to greater
potential for holistic reconstruction strategies. The minimum-ionising-particle (MIP) timing 
detector (MTD) is divided into a barrel section with an acceptance of $\eta < 1.5$ and an 
endcap section with an acceptance of $1.6 < \eta < 3.0$. The silicon part of the 
high-granularity endcap calorimeter (HGCAL), which supports high-precision timing 
measurements, covers an acceptance of $1.6 < \eta < 3.0$. While neither detector extends 
as far forward as the HGTD, they enable precision timing measurements across a wide kinematic
range for both charged and neutral particles in the endcap region, and for charged particles 
and converted photons in the barrel. In addition, the CMS barrel calorimeter, which has an 
acceptance of $\eta < 1.5$, will enable precision timing measurements for highly energetic 
neutral particles in the barrel region.

Particles traversing the MTD acceptance traverse less material on average than those 
traversing the HGTD acceptance, and the material distribution in front of the MTD is generally
more uniform than in front of the HGTD. As a result the MTD has a significantly higher efficiency,
around 80\%, to associate time information to pions, and this efficiency is essentially uniform
in both kinematics and geometry apart from a gap in $1.5<\eta<1.6$ in the detector coverage.
On the other hand, periodic dead areas in the detector acceptance mean that the efficiency to
associate time information to muons is only $90-95$\% in the barrel region, while it rises to
100\% in the endcap region. The high efficiency for hadronic particles and the wide geometric 
acceptance also mean that the MTD is able to associate timing information to nearly all 
collisions of interest, in turn enabling a very general use of particle-collision matching
in physics analysis.

\begin{figure}[t]
    \centering
    \includegraphics[width=0.9\linewidth]{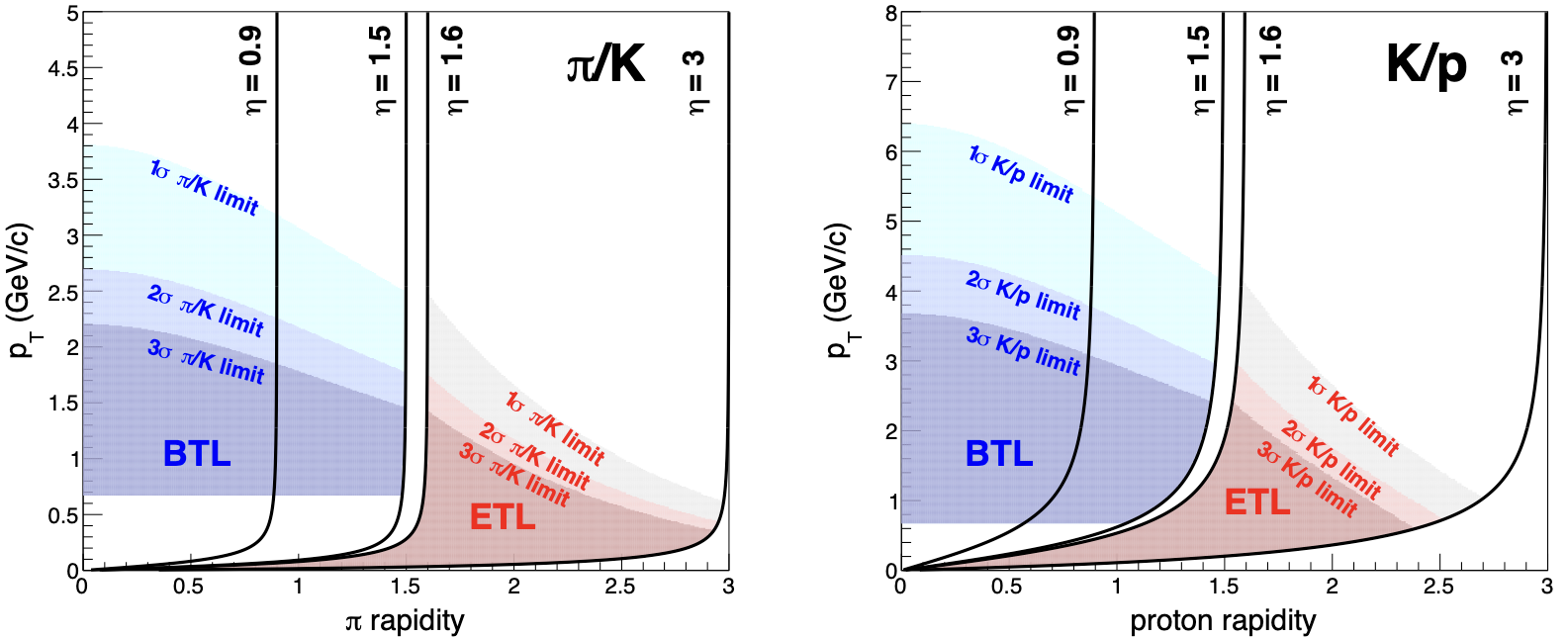}
    \caption{Expected performance for charged particle identification in $p_T$ and rapidity in an early 
    run of the HL-LHC with the proposed MTD, with a time resolution of about 30 ps, which is achievable 
    since there will not yet be significant radiation damage. Different colors of the shaded regions 
    correspond to 1, 2 and 3$sigma$ separations. Left: The $pi/K$ separation vs $p_T$ and rapidity $y$ 
    under the pion hypothesis. Right: The $K/p$ separation vs $p_T$ and rapidity $y$ under the proton 
    hypothesis. Shown also are contours in pseudo-rapidity, $|\eta|$. The label BTL in these plots 
    indicates the barrel timing detector and the label ETL indicates the endcap timing detectors. 
    Caption and figure reproduced from~\cite{CMS:2667167} as permitted by the article's licence.}
    \label{fig:cms-mtd-tof-perf}
\end{figure}

One example of a qualitatively different use of timing information which is possible in CMS
is particle identification through time-of-flight information. 
While time-of-flight measurements would require femtosecond time resolution for typical LHC
particle momenta, it is possible to use the MTD as a time-of-flight particle identification
detector in the $p<4-5$~GeV and $\eta < 2$ range as shown in Fig.~\ref{fig:cms-mtd-tof-perf}. 
This also requires the time of the original collision to be well-measured, but the combination
of central and forward coverage of the MTD makes that significantly more efficient in CMS 
than in ATLAS. While primarily referenced in the context of $PbPb$ collisions in the MTD
technical design report, it is possible that the MTD could be useful as a particle identification
detector for heavy-flavour signals also in $pp$ collisions.

Another example is the reconstruction of beyond Standard Model long-lived particles (LLP), which often
decay in cascades involving a substantial amount of missing energy. Because of the combination of missing
energy and a highly displaced decay vertex, it is not possible to achieve a useful mass resolution on the
LLP using classical missing energy techniques. The MTD can however be used to measure
the time of their visible decay products and in turn the decay vertex itself. The time difference 
between the decay vertex and the production vertex can then be interpreted in terms of the LLP's 
velocity and, when combined with the kinematics of its visible decay products, the LLP's mass. 
The MTD can also help identify LLP decay products by looking for particles whose arrival time is
significantly later than that of relativistic Standard Model particles produced directly in the 
$pp$ collision, and the combination of these methods will help to maintain or improve the CMS detector's
sensitivity to a wide range of LLP signatures despite the much higher HL-LHC pileup. 

Beyond these examples, benefits of the MTD detector's ability to suppress pileup are expected in 
track isolation, jet flavour tagging, and the identification of the origin vertices of signal 
candidates across a wide range of putative signals. Similarly, the HGCAL's precision timing, including
for low-energy neutral objects, is expected to play a significant role in suppressing pileup and
thus improving the resolution of endcap calorimeter objects in the HL-LHC period. The availability
of timing information for charged particles in almost the full detector acceptance may, in the long
run, also allow timing to be used in order to remove pileup at the hit level and improve tracking 
performance for hard-to-reconstruct tracks associated with $pp$ collisions of interest. We will
now explore this possibility in more detail with reference to the proposed second upgrade of the
LHCb detector.

\subsubsection{Timing use-cases in LHCb}

While ATLAS and CMS are hermetic detectors, LHCb is a forward spectrometer with an approximate 
acceptance of $2 < \eta < 5$ and a tracking system split around a dipole magnet. It currently (Upgrade~I) 
operates at a pileup around 10 times smaller than that of ATLAS and CMS, and its proposed HL-LHC 
upgrade (Upgrade~II) will operate with a pileup between 5 and 7.5 times smaller than that of ATLAS 
and CMS depending on the precise design and operating scenarios of the three detectors. In contrast 
to ATLAS and CMS, where the full-granularity timing information is only guaranteed to be available 
after a positive first-level trigger decision, LHCb will maintain its current full detector readout 
in the HL-LHC period and will consequently also have access to precision time information at the first 
level trigger. As we shall see, this opens up the possibility of making precision timing a first-class 
citizen in both the detector reconstruction and selection logic. 

\begin{figure}[t]
    \centering
    \includegraphics[width=0.7\linewidth]{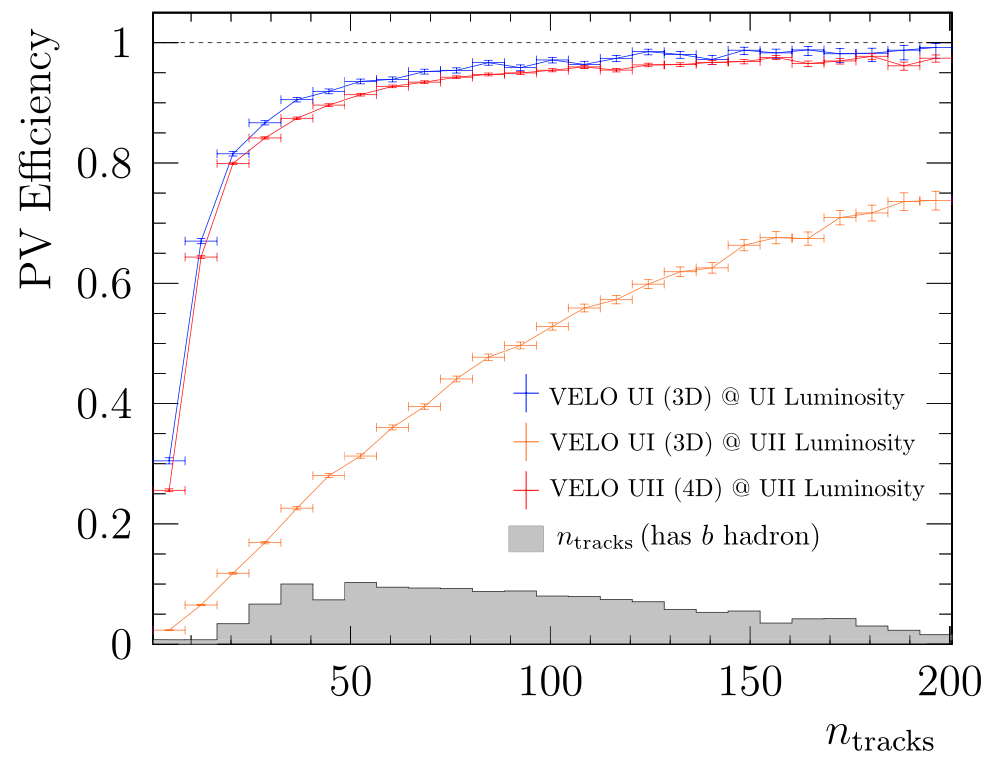}
    \caption{Collision vertex (PV) reconstruction efficiency as a function of track multiplicity for the 
    VELO Upgrade~II Baseline scenario, and for the current (Upgrade I) VELO in both 
    Run~3 and Upgrade~II conditions. Caption and figure reproduced from~\cite{LHCbcollaboration:2903094} 
    as permitted by the article's licence.}
    \label{fig:lhcb-u2-timing-pv}
\end{figure}

Because the LHCb vertex detector (VELO) is located outside the experiment's dipole magnet, vertex detector
tracks are straight lines and the efficiency with which they can be reconstructed by a pixel detector of
reasonable spatial granularity is close to 100\% in the full geometric and kinematic detector acceptance. 
The availability of precision timing information in the vertex detector will therefore allow all tracks 
originating in the interaction region and falling within the LHCb detector acceptance to be timestamped, 
in turn allowing all reconstructed collision vertices to be timestamped and the entire bunch crossing image 
formed within the vertex detector to be projected out in both space and time. As illustrated in 
Fig.~\ref{fig:lhcb-u2-timing-pv}, this allows LHCb to retain its current $pp$ collision reconstruction 
efficiency in Upgrade~II, despite the much higher pileup. Notice that LHCb's vertex detector covers a 
similar acceptance as ATLAS's HGTD detector, but the integration of the timing and spatial information
inside the pixel tracker avoids material interaction inefficiencies inherent to a post-tracker timing
layer, and allows a greater number of hits containing timing information which in turn leads to a better
overall time resolution for both tracks and $pp$ collision vertices.

\begin{figure}
    \centering
    \includegraphics[width=0.7\linewidth]{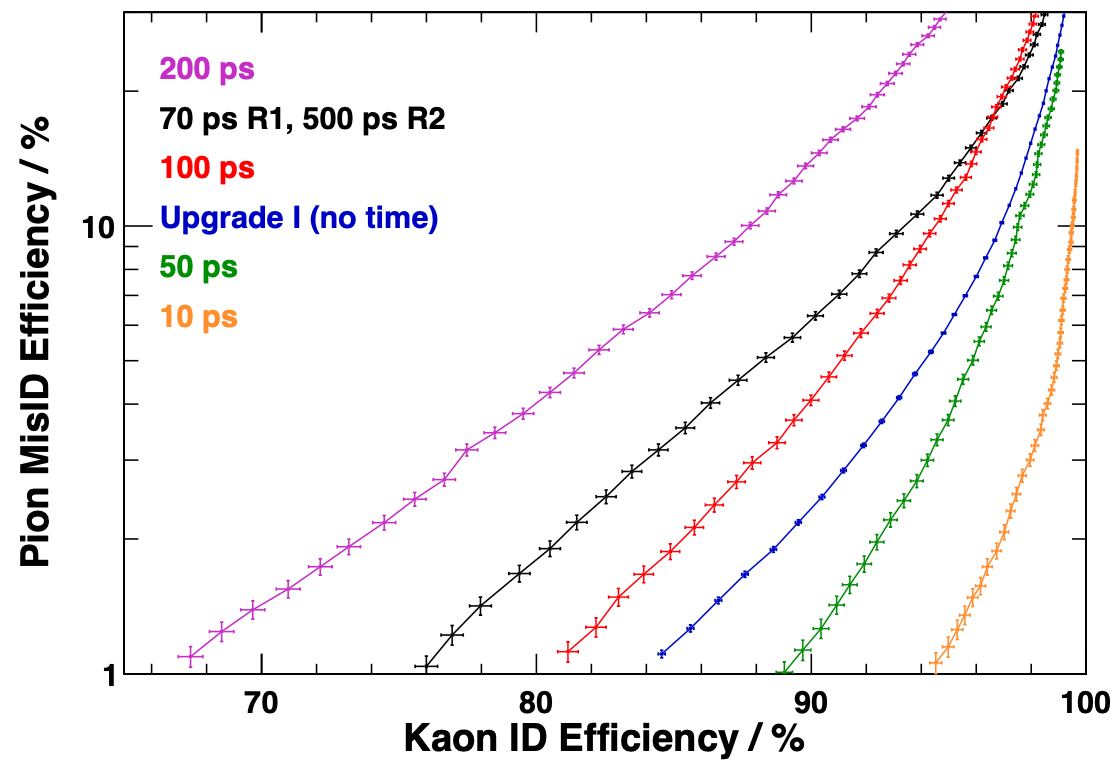}
    \caption{LHCb Upgrade~II RICH performance to discriminate between pions and kaons, plotted with 
    different time gates applied to the detector hits. The contour 
    marked ‘Upgrade I’ corresponds to the performance at the lower luminosity of $2e33$~cm$^{-2}$s$^{-1}$ 
    and with no time gate. Figure and part of the caption reproduced from~\cite{LHCb:2021glh} as 
    permitted by the article's licence.}
    \label{fig:lhcb-u2-timing-rich}
\end{figure}

Precise identification of charged hadron species across a wide ($2-100$~GeV) kinematic acceptance is
a key capability of the LHCb detector and enables much of its heavy flavour physics programme. While
LHCb uses neural-network classifiers based on information from the whole detector to achieve the best
particle identification performance, most of its disciminating power for charged hadron species comes
from information provided by two ring-imaging Cherenkov (RICH) detectors: RICH1, which is located 
between the VELO and dipole magnet, and RICH2 which is located between the downstream tracker stations 
(MT) and the electromagnetic calorimeter (ECAL). As with the VELO, the integration of timing and
spatial information within a single detector means that all tracks which generate Cherenkov light
within the RICH detector acceptances can be projected in time and space, mitigating the impact of the
increased detector occupancy at the higher pileup of Upgrade~II. One way to think about this is in
terms of ``time gates'' which can be applied to the RICH detector hits in order to select those which
are compatible in time with a specific $pp$ collision reconstructed and timestamped by the VELO. The 
width of the time gate depends on both the intrinsic RICH and VELO time resolutions, but in practice 
the RICH resolution dominates since the VELO is able to reconstruct collision vertices with time 
resolutions below 10~ps in most cases. The impact of time gating is illustrated in 
Fig.~\ref{fig:lhcb-u2-timing-rich} as a function of the assumed time gate with. The current 
best estimate~\cite{LHCbcollaboration:2903094} is that LHCb will be able to build an upgraded RICH 
system capable of efficiently applying time gates of $150-300$~ps in width, but the technology 
R\&D is still ongoing.

Similar performance arguments apply to LHCb's ECAL, which will be upgraded to both greatly improve its
spatial granularity and to introduce precision timing information across a wide range of cluster
energies of interest to LHCb. The downstream part of LHCb's tracker, which is positioned between the 
magnet and the RICH2 detector, currently consists of 2.5~metre long Scintillating Fibre (SciFi) cables 
arranged in 12 alternating vertical ($x$) and angled (stereo) layers to allow a 3D track reconstruction, read
out by sillicon photomulipliers cooled to $-40^\circ$~Celsius. Its occupancy would become unacceptably
high in Upgrade~II conditions, and so LHCb plans to introduce a CMOS-based pixel tracker into the
detector region nearest the beampipe, while keeping the SciFi detector in the outer regions for cost
reasons. The CMOS tracker has a sufficient spatial granularity that it is not necessary to equip it
with precision timing information in order to ensure a good track reconstruction efficiency in Upgrade~II
conditions. In addition, since most tracks of interest to physics analysis have a VELO segment, the
CMOS tracker segments can be timestamped from their matching VELO segment and can therefore be projected
in time in order to reduce pileup backgrounds at the analysis level. In the case of the SciFi, it may
be possible to introduce modest timing capabilities, between $0.8-1$~ns per hit, without incurring
significant additional costs. Because the SciFi layers rely on the $x$-stereo layout for a 3D track
reconstruction, their reconstruction suffers from large combinatorial backgrounds even once the SciFi 
occupancy has been reduced by the inner CMOS tracker. This combinatorial burden may be reduced by using
timing information directly in the track reconstruction, and work is ongoing to quantify the precise
benefits of such an approach.

\subsection{Potential holistic uses of timing information}
Up to this point the examples of how timing information is used have been fairly atomic: precision
timing measurements in a given detector help suppress pileup backgrounds in a specific signal
selection or reconstruction step. Even in cases like time-of-flight particle identification, which
requires both the origin and endpoint of the particle's trajectory to be timestamped, or NA62's 
coincidence matching between the Gigatracker and other timing and particle identification detectors,
the timing information does not shape how the overall detector reconstruction is performed. But once
a large enough fraction of a detector can be precisely projected out in four dimensions, it becomes
possible to envisage a much more intimate connection between space and time information in the
data processing and the building of high-level physics analysis objects. Let's illustrate this by 
reference to the LHCb Upgrade~II detector and its real-time analysis paradigm.

As discussed in the introduction and shown in Fig.~\ref{fig:lhcbbbarvspileup}, at high pileups the
fraction of bunch crossings which contain interesting physics processes far exceeds the available 
permanent storage. The real-time analysis concept addresses this by saving only information directly
relevant to any given physics analysis, typically consisting of information associated with a single
initial collision as well some selected information about the overall event to help with background
and systematics studies, to permanent storage. The limitation of this approach in a purely spatial 
detector is that it requires reconstructed objects to be associated with a collision, so it cannot 
be used to classify individual detector hits or reconstructed objects from long-lived particles whose 
spatial association to an initial collision is often ambiguous. Precision timing information lifts
both of these limitations.

Consider a typical displaced vertex trigger selection, which has been the mainstay of flavour physics
triggers at hadron colliders since CDF.~\cite{CDF:2003mka} This trigger typically selects pairs of tracks
forming a vertex displaced from the initial collision, whose kinematics are compatible with those of
a beauty or charm hadron decay. At high pileup, this trigger cannot effectively reduce the rate of
bunch crossings, since too many bunch crossings contain such decays. Within any given bunch crossing,
however, a displaced vertex trigger can effectively select a small number of interesting initial collisions
so long as the detector occupancy remains low enough such that fake tracks do not dominate the
reconstruction output. In the case of LHCb specifically, it is already clear based on information
documented in~\cite{LHCbcollaboration:2903094} that it will be possible to efficiently reconstruct 
two-body displaced vertices from beauty and charm decays at 30~MHz in Upgrade~II pileup conditions 
no matter what is assumed about the detector budget and technology evolution. Because these displaced
vertices can be timestamped by the vertex detector, they can also be efficiently associated to the 
initial $pp$ collisions. This in turn provides a time gate for the rest of the detector, 
similarly to the logic already discussed in terms of the RICH particle identification. 

Since the interaction region has a three-sigma length in time of around $\pm 600$~ps, and a reconstructed
$pp$ collision vertex typically consists of at least 4 tracks with a time resolution of 20~ps each, 
each interesting $pp$ collision vertex selects a time gate corresponding to $\lesssim\! 2$\% of the total
interaction region. Whether other detector hits or reconstructed objects fall within this time gate
will therefore be governed by their own intrinsic resolution. One possibility which emerges from this
kind of gating trigger is to extrapolate timestamped tracks into detectors which do not provide precision
timing measurements and use track-to-hit matching to partially time-gate them. For example, LHCb could
first reconstruct relatively high-momentum tracks which are relevant for the secondary vertex 
selection, extrapolate them to the downstream tracker and calorimeter, and use them to remove hits
which are clearly incompatible with the relevant time gate. A lower momentum reconstruction could then
be run at the effectively reduced detector occupancy, the same time gating performed, the detector
occupancy further reduced, and so on. In effect, the time information allows to transform a situation
where the collision of interest is a minority in a bunch crossing dominated by pileup, to one in
which the residual in-time pileup is a minority in a time-gated event dominated by the collision of 
interest.

This is particularly interesting because while the computational cost to reconstruct a detector increases 
at least linearly (and often worse than linearly) with occupancy, tracks which are of interest to inclusive
triggers tend to have higher momenta, follow straighter lines, and be inherently easier to reconstruct. 
Such time-gating may therefore enable a more complete detector reconstruction at higher rates, closer to 
the full collider bunch-crossing frequency. The eventual utility of such an approach depends on the fraction 
of the detector which provides precision timing, but it is not an all-or-nothing proposition: time-gating 
can be applied to a subset of trigger paths and the rate-efficiency working point smoothly varied depending 
on the physics requirements as with any other online selection. While this example was illustrated with 
respect to LHCb, it might also be relevant for CMS, albeit only at the output rate of its hardware trigger, 
and for other future detectors with precision timing in the full (or nearly full) detector acceptance.

\section{Overcoming network limitations in four-dimensional detectors}
While precision timing information is certainly a powerful tool, it also significantly increases
the data rates produced by a HEP detector. This can be illustrated with reference to a zero-suppressed
subdetector whose data rate scales linearly with pileup (once the pileup itself is in the Gaussian rather
than Poissonian regime) such as the LHCb vertex detector. The estimated data rate of LHCb's
Upgrade~II timing vertex detector is 34~Tbit/s, around twelve times that of the LHCb
Upgrade~I vertex detector despite operating at only a $5-7.5$ times higher pileup. 
Similar considerations apply to other timing detectors, and constitute a major challenge
for the future readout of HEP detectors.

\begin{figure}[t]
    \centering
    \includegraphics[width=0.7\linewidth]{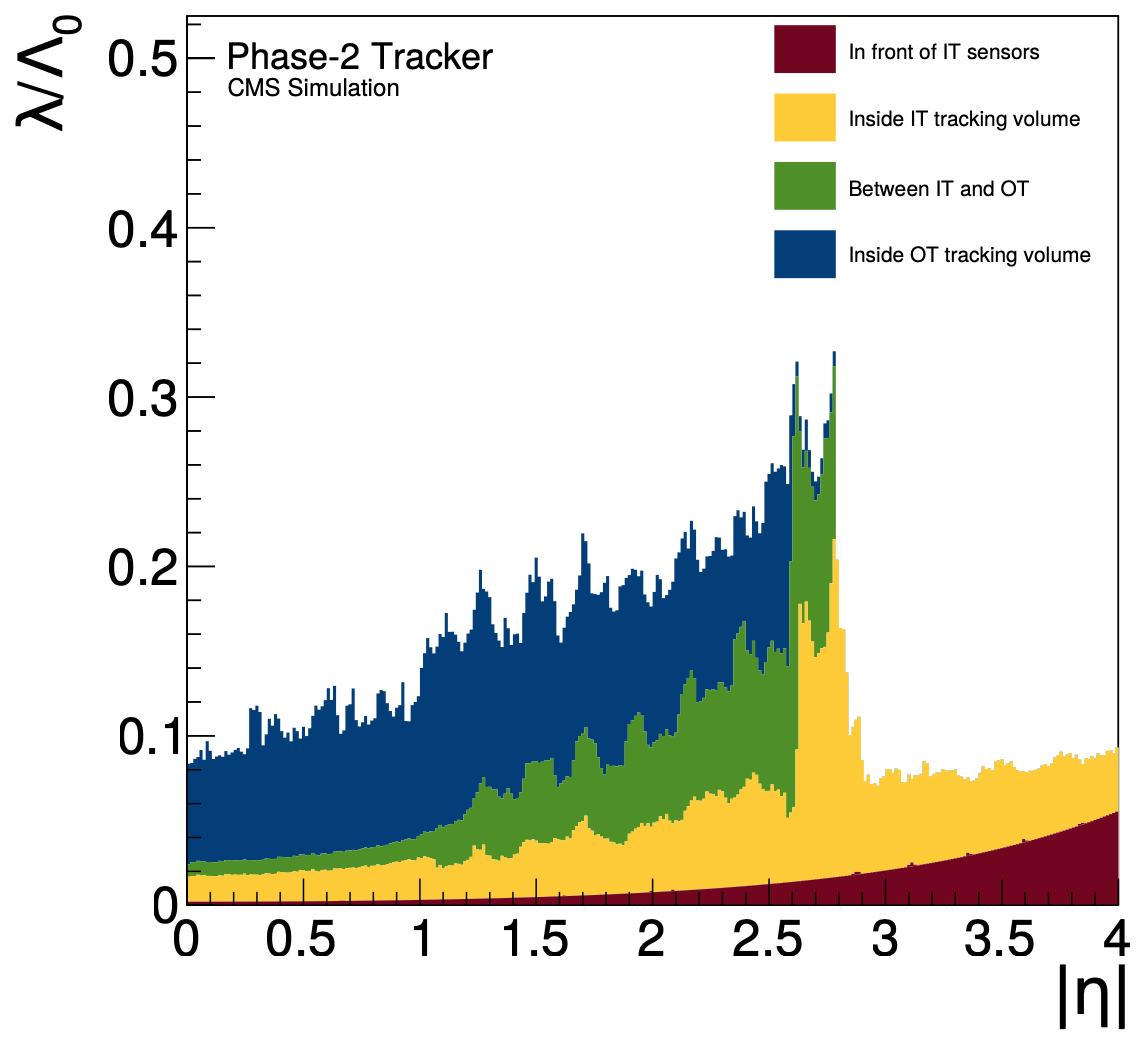}
    \caption{Material budget inside the tracking volume estimated in units of nuclear interaction 
    lengths of the Phase-2 CMS detector. The material in front of the Inner Tracker sensors is 
    shown in brown, that inside the Inner Tracker tracking volume in yellow, that between IT and 
    OT sensors in green, and that inside the Outer Tracker tracking volume in blue. The histograms 
    are stacked. Figure and caption reproduced from~\cite{CERN-LHCC-2017-009} as 
    permitted by the article's licence.}
    \label{fig:cms-p2-tracker-material}
\end{figure}

While data rates are primarily a cost driver for detectors
with an open geometry such as LHCb, since the readout links can be placed outside the detector's 
acceptance, they are among the most important constraints which limit the rate at which the 
innermost subsystems of hermetic detectors can be read out. This is illustrated for the case of
the CMS HL-LHC (``Phase-2'') detector in Fig.~\ref{fig:cms-p2-tracker-material}, which shows the 
material budget of its tracker. The yellow inner tracker volume represents about 25\% of the total
material budget in most of the central ($\eta \lesssim\! 2.5$) detector region. The readout
links required by the inner tracker contribute a significant fraction of this material budget,
between about $0.48-1.96$~grams per metre of cable, depending on the inner tracker module in 
question, compared to $2.33-3.75$~grams for the modules themselves and other associated services.
This comparison does not consider the electrical power and the required cooling associated with the detector readout, but is 
already sufficient to understand why the CMS inner tracker is only read out at 1~MHz and not the 
full LHC bunch crossing frequency. Doing so even at some reduced granularity would significantly 
increase the overall tracker material budget and power consumption and therefore degrade the 
detector's performance. 

The speed at which radiation-hard links can transfer data is both significantly slower than the 
fastest commercial Ethernet standards and is limited by transceiver technology, rather than by 
the cables. The lpGBT link standard~\cite{10778249} developed for HL-LHC applications enables
low-power 10.24~Gb/s data transfers from the detector, whereas 800~Gb/s Ethernet was approved by
the IEEE in 2024. In addition,
\href{https://www.nngroup.com/articles/law-of-bandwidth/}{Nielsen's law} continues to hold in 
consumer applications, although with something of a recent slowdown compared to the most optimistic
predictions~\cite{DBLP:journals/corr/TrotmanZ13} made a decade ago. Nevertheless it is clear that
if the material budget (and power consumption) considerations can be addressed, networking cost will 
not prevent the full readout of future HEP detectors even if they deploy precision timing throughout.

This assertion is borne out by the evolution of the LHCb experiment's detector readout, from the
first LHCb detector~\cite{CERN-LHCC-98-004} which operated a hardware trigger and a 1~MHz detector
readout, to the current Upgrade~I detector~\cite{Bediaga:1443882} which is fully read out, and towards
the proposed Upgrade~II detector. The current costing~\cite{LHCbcollaboration:2903094} 
of LHCb Upgrade~II, which will transfer an estimated 170~Tbit/s off the detector across
around thirty thousand lpGBT links, contains a network link cost estimated at $4.7-7$~MCHF depending on
the chosen pileup working point. This is only around 5\% of the overall estimated detector cost,
and significantly smaller than the estimated cost of both the FPGA boards required to receive 
this data and the computing power required to process it. It is also very similar, when adjusted
for inflation, to the cost of the optical links required to read out the current (Upgrade~I) LHCb 
detector as estimated at the same point in the project's lifecycle. Further support for an optimistic
outlook on networking costs is provided by the Phase~2 CMS upgrade, which will read out all its subsystems 
except the innermost tracker at the full LHC bunch crossing rate in order to make their information available
to CMS's Phase~2 first-level trigger~\cite{CERN-LHCC-2020-004}, albeit in some cases with reduced granularity 
and without precision timing information. It is therefore safe to assume
that the next generation of collider experiments, which are only due to come online in the 2040s or
2050s, will present a tractable network bandwidth problem.

In order to address the material and power budget limitations which currently prevent the full readout
of the innermost subdetectors in hermetic collider experiments, it is necessary to develop
radiation hard transceivers which can support a roughly one order of magnitude greater data rate. 
Silicon Photonics~\cite{Prousalidi:2024lbl} (SiPh) is a promising technology in this regard for a number
of reasons, not least of which is the fact that it shares the same underlying silicon technology
and often the same fabrication processes as the silicon sensors which are now the standard for
innermost collider experiment subdetectors. This allows a much closer integration of the 
transceiver and sensitive detector elements, simultaneously minimizing power consumption, required
detector services, and thereby material costs. Silicon photonics requires the use of single-mode optical
fibres, which also naturally leads to longer reach. Recent R\&D studies~\cite{Scarcella:2025rdl} have
demonstrated that data rates of 25.6~Gb/s are achievable so long as the lengths of the wire bonds
which connect the photonic integrated circuits to the transmitter ASICs remain sufficiently short, 
$<800$~$\mu$m for the system in question. Wavelength division multiplexing can then be used to
achieve 100~Gb/s transmission speeds, roughly the one order of magnitude improvement over the current
lbGBT links which is required. A particular advantage of silicon photonics compared to current
optical readout systems is that the laser sources used to convert electrical signals to optical ones 
are placed off-detector. This both reduces the material budget inside the detector, which we have
seen is critical to the viability of a full detector readout, as well as reducing the power and cooling
requirements. It also means that the laser sources are placed in (largely) radiation-free zones. 
While the distant placement of these laser sources brings its own challenges, in particular with
managing the polarisation of the light provided by the laser source~\cite{10234459}, these appear
to be tractable, particularly when considering the timescales associated with post-HL-LHC experiments.

\section{Real-time four-dimensional processing beyond the HL-LHC}

As can be seen from Fig.~\ref{fig:data_rates_cerri}, the HL-LHC detectors, and LHCb Upgrade~II in
particular, will represent the most significant real-time processing challenges in HEP for some time.
Both the planned LHCb and ALICE HL-LHC upgrades (``Run 5'' in the figure) are building on proven 
full detector readouts which the collaborations deployed during Run~3, and there is no reason to doubt
that their basic processing model will scale as required. While the price of COTS computing technologies
is extremely volatile right now, and subject to increasing geopolitical tensions, the underlying hardware
continues to improve in performance whether we are discussing CPU, GPU, or FPGA processors. And after a 
decade of painfully transitioning legacy software stacks consisting of millions of lines of code to enable
parallel and heterogeneous data processing, the next generation of experiments is well-positioned to reap
the rewards. This is true both of the large LHC collaborations, who have the personnel to support their
own processing frameworks and approaches to parallelism, and of upcoming smaller experiments which will
benefit from CERN's support for the development of the Key4HEP~\cite{Key4hep:2023nmr} software stack as
a turnkey foundation on which to build. For example the ePIC~\cite{dalla_torre_2026_19496158} detector 
at the planned Electron-Ion collider plans for a full detector readout and real-time processing, relying
on a carefully calibrated zero-suppression of the individual subdetectors in order to manage the data
rates produced by its high granularity but very low occupancy subsystems. The case studies considered
here are chosen to illustrate certain specific challenges facing proposed future detectors and facilities
in their real-time data processing.

\subsection{Viability in FCC-ee detectors}

The online computing challenges at the FCC have been discussed in~\cite{Brenner:2021mxb} and more recently
a publicly-available \href{https://indico.cern.ch/event/1529896/contributions/6436830/attachments/3036746/5363075/FCCee_TDAQ_EoI-200325.pdf}{expression of interest} for which no published record unfortunately exists. The essential challenge
confronted by the FCC-ee experiments is not the signal rate, which is at most around 200~kHz when running
at the $Z$ pole (TeraZ), but rather the challenge of reading out the background caused by incoherent
pair production across all 50~MHz of beam crossings without increasing the detector's material budget
in a way which would compromise the physics sensitivity. In addition, any real-time processing inefficiencies
must be understood at the $1e^{-4}-1e^{-5}$ level in order to achieve the sensitivity on the recorded
luminosity necessary for the precision cross-section measurements. 

These challenges nevertheless appear tractable in the context of detectors which will produce relatively 
small data volumes and do so two decades from now. The most extreme example is perhaps the IDEA DR 
Calorimeter, whose granularity and high-intrinsic-noise design leads to a background data rate of 1.6~TB/s 
which is around two orders of magnitude greater than the signal data rate. While seemingly large, this 
rate is smaller than that of today's LHCb Upgrade~I detector and an order of magnitude smaller than the 
anticipated data rate of the LHCb Upgrade~II detector. It seems unlikely to present an insurmountable 
challenge a decade after LHCb Upgrade~II is planned to go into operation. Note that while picosecond timing
is not essential for differentiating bunch crossings in the baseline FCC-ee configuration, with LHC-like
25~ns bunch spacing, significantly shorter bunch spacings, down to 1.25~ns, are also under investigation 
and would change this calculus. In addition, even at the baseline bunch spacing, certain FCC-ee detector
concepts such as the ILD~\cite{ILD:2025yhd} are considering picosecond timing layers in the tracker or
calorimeters, whether for time-of-flight particle identification or neutral particle tagging. The addition
of such timing information will increase the material and power budget constraints on the detectors,
but will not meaningfully change the conclusion that the overall data rates will be small compared to 
those of the HL-LHC experiments.

It is also worth addressing the worries about network reliability in the context of a full detector 
readout expressed in~\cite{Brenner:2021mxb}. No detector readout can fully eliminate dead time,
but in practice the current LHCb detector has demonstrated steady-state operations with a few 
permille of dead time, much of which is caused by the loading of calibration constants near the
start of a fill and could be further reduced if necessary by employing buffers to hold data while
the constants are loaded. Moreover dead time is not an obstacle to a precise luminosity measurement
in and of itself: what matters is that the subset of randomly
selected events used for measuring the luminosity have an equal probability to suffer dead time as
physics events. This effectively amounts to a requirement that dead time should not be caused by
an inability to read out or process particularly large events. While this is difficult to avoid in
general during real-time processing, it is in some sense a more difficult problem for triggered
experiments whose fixed-latency first level processing has a strict cutoff on the computing time
available for each event. A triggerless real-time processing is by contrast only concerned about
keeping the average throughput above a certain value, and only runs into trouble if events are so
large as to exceed the available processor memory. In any case, there is no fundamental limitation 
on how precisely network packet losses can be monitored.

Of course, the FCC-ee experiments may choose to not implement a triggerless readout because they
don't necessarily need one for the physics, or because the benefits of having the smallest possible
material budget outweighs the convenience of a full detector readout. With a signal rate\footnote{While
the total event rate can be relevant for estimating the detector readout complexity, the signal
rate is the most relevant for estimating the real-time processing complexity, as non-signal events
can typically be identified and removed from consideration in a fraction of the overall processing time.} 
of at most 200~kHz, a factor seven smaller than that processed by LHCb's physics-quality reconstruction 
today, and events which are far simpler than those of the HL-LHC detectors, 
this convenience should not be underestimated. If this factor seven is combined with a cost-performance 
improvement in COTS computing of $10\%$ per year, significantly less than estimates traditionally
used~\cite{Butler:2055167,CERN-LHCC-2015-020,LHCbcollaboration:2903094} by CERN experiments, 
the overall cost to perform a physics-quality real-time processing of an FCC-ee experiment will be 
about 70 times smaller than that of LHCb today. Another way to look at this is that compared to the
roughly 3500 CPU servers which LHCb currently uses, the FCC-ee experiments will need around 50 servers
with the technology of the 2040s. Such a compact system will also be significantly easier to operate, and
therefore have inherently fewer difficulties with network scaling, than the equivalent systems of
the HL-LHC experiments.

\subsection{Viability in muon collider and FCC-hh detectors}

The topic of muon collider experiment data rates was recently reviewed in~\cite{Holmes:2025ybv}, and
many of the same arguments made in respect of FCC-ee apply here as well. For the 10~km collider option
the bunch crossing rate is around 30~kHz and the overall detector
rate is dominated by backgrounds from the decays of beam muons, with
signal processes of interest occurring several orders of magnitude below the bunch crossing rate.
All this is significantly below the rates processed by the HL-LHC detectors, even assuming highly
granular and fully four-dimensional detectors. It is worth stressing that while four-dimensional
trackers or calorimeters are something of a luxury at FCC-ee, and would primarily serve to enhance
sensitivity to exotic long-lived particles, picosecond timing detectors are much more obviously 
useful in order to optimally suppress muon beam backgrounds at the muon collider.

The current FCC-hh design~\cite{Schulte:2025pgs} calls for the same 25~ns bunch spacing as used at 
the LHC and HL-LHC, but a nominal pileup of 1000, compared to a peak pileup of around 200 expected 
at ATLAS and CMS in the HL-LHC period. At such a pileup, any hope of operating a general-purpose
detector rests on the use of precision timing information to separate the pileup collisions.
Considering that the detectors in question will operate in the 2070s or later, a decade further
removed from today than the LEP experiments were from the founding of CERN, any statements about
detector readout or computing are more than speculative. If we do not let this deter us, we can
note that the same cost-performance improvement in COTS computing of $10\%$ per year assumed in our
FCC-ee discussion would lead to two orders of magnitude cheaper data processing by the 2070s. 
Even assuming that the data processing complexity scales quadratically with pileup, which was the
case for the original LHC experiments, the expected COTS improvement outstrips the increased pileup
by factors. And the problem of processing scaling with pileup is being widely 
addressed~\cite{CERN-LHCC-2017-020,Collaboration:2759072} in the runup to the HL-LHC, so there are
good reasons to think that this is a pessimistic assumption. If we are more optimistic and assume
that the data processing complexity scales linearly, then the COTS cost-performance improvements
(factor 100) nearly cancel the complexity of the additional pileup (factor 5) multiplied by the 
increase in processing rate if the FCC-hh detectors are read out at the full collider bunch crossing
rate (factor 30). And such extrapolations are based on a simple scaling of existing, proven, technology,
while fully ignoring the coming five decades of methodological improvements or entirely novel
processing technologies which may become commerically viable. The back of the envelope estimates,
therefore, favour optimism.

\section{Conclusion and non-HEP uses}
Over the past four decades, high-energy-physics experiments have seen their data rates grow from around
ten megabytes per second to the tens of terabytes per second expected at the high-luminosity LHC. 
Over the same period commercial networking, processing, and storage technologies have kept pace
with developments in HEP, and the data processing problems faced by HEP experiments have proven
remarkably well-suited to modern parallel processing architectures. Because of this, today's HEP 
experiments process more of their data in (quasi)-real-time than ever before, and do so with a 
fidelity that is closer than ever, and in some cases identical, to that required by physics analysis. 

There is no reason to expect computing technology to stop evolving in the coming decades, while
HEP experiments face a period of constant or decreased data rates between the HL-LHC and the
proposed FCC-hh colliders. Ongoing technological developments in radiation-hard data links,
pioneered by high-energy physicists and CERN, hold the promise of low-mass and low-power 
readouts reaching into the innermost regions of hermetic detectors. At the same time, HEP detectors
are increasingly adding precision timing information as well as improving spatial granularity,
in order to both improve their signal-to-background discrimination and enable qualitatively 
different observables to be measured. The confluence of these factors is of interest far beyond
HEP, in particular because HEP detectors continue to generate and process some of the largest 
data rates on earth. While sensors capable of managing such data rates while providing picosecond 
timing, micrometer resolution, and reliable operation in extreme radiation environments may 
not have widespread commercial applications today, their very existence may provoke such 
commercial applications to be developed.

In order to make the most of these opportunities, it is essential that the field maintains an
optimistic outlook focused on the long-term trend towards four-dimensional detectors which can 
be fully, or nearly fully, processed in real-time with a physics-analysis quality alignment, 
reconstruction, and calibration. Such methods may not be strictly necessary in some of the 
facilities which will bridge the gap between HL-LHC and FCC-hh, but with that gap stretching 
beyond three decades there is also no guarantee that the field can retain the relevant technical 
know-how if ``good enough'' approaches are adopted at FCC-ee, the EIC, and other contemporary 
facilities. If on the contrary we seize the opportunity to rethink our detectors as fundamentally
four-dimensional objects, we may yet find that the holistic benefits of such an approach far exceed
the expected improvements in any single area. This review has, I hope, served to make the case
that such optimism is justified and timely.

\section*{Acknowledgements}
I would like to thank Niko Neufeld, Renato Quagliani, and Zekun Jia for useful discussions which informed
this article.
The manuscript was researched and written without any use of artificial intelligence, automated copy-editing 
tools, or writing assistants. All errors, omissions, and other shortcomings are my own. 

\bibliographystyle{unsrtnat}
\bibliography{my-bib-database}

\end{document}